\documentstyle[emulateapj,psfig]{article}

\begin{document}


\title{Pulsar braking index: A test of emission models?}

\author{R.X. Xu$^{1,2}$ \& G.J. Qiao$^2$}

\altaffiltext{1}{National Astronomical Observatories, Chinese
Academy of Science, Beijing 100012, China}

\altaffiltext{2}{Astronomy Department, Peking University, Beijing
100871, China}

\begin{abstract}

Pulsar braking torques due to magnetodipole radiation and to
unipolar generator are considered, which results in braking index
being less than 3 and could be employed to test the emission
models.
Improved equations to obtain pulsar braking index and magnetic
field are presented if we deem that the rotation energy loss rate
equals to the sum of the dipole radiation energy loss rate and
that of relativistic particles powered by unipolar generator.
The magnetic field calculated by conventional way could be good
enough but only modified by a factor of $\sim 0.6$ at most.
%
%
Both inner and outer gaps may coexist in the magnetosphere of the
Vela pulsar.


\end{abstract}

\section{Introduction}

Pulsar emission process is still poorly understood even more than
30 yr after the discovery.
Nevertheless, it is the consensus of the researchers (e.g., Usov
2000) that primary pairs are produced and accelerated in regions
(gaps) with strong electric field along magnetic line
($E_\parallel$) and more secondary pairs (with multiplicity $\sim
10^{2\sim 4}$) are created outside the gaps ($E_\parallel=0$), and
that instability may be developed in the secondary $e^\pm$
relativistic plasma in order to give out coherent radio emission.
Numerous models have been made concerning gap acceleration,
whereas it is urgent to find effective way to test those specific
and detailed models against observations.

Because of observational difficulties, only braking indices
$n\equiv {\Omega \ddot \Omega/ \dot{ \Omega}^2}$ ($\Omega$ the
angular velocity of rotation) of 5 young radio pulsars have been
obtained observationally (Lyne \& Graham-Smith 1998 and
references therein, Camilo et al. 2000). They are: %
PSR B0531+21 ($n=2.51\pm 0.01$), %
PSR B1509-58 ($n=2.837\pm 0.001$), %
PSR B0540-69 ($n=2.2\pm 0.1$), %
PSR B0833-45 ($n=1.4\pm 0.2$), %
PSR J1119-6127 ($n=2.91\pm 0.05$). %
Certainly, these observed indices include precious information on
how pulsars produce radiation. But all of them are remarkably
smaller than the value $n=3$ expected for the pure magnetodipole
radiation,
%
according to which, the polar magnetic field strength at pulsar
surface, $B$, is conventionally determined by (e.g., Manchester \&
Taylor 1977),
\begin{equation}
B={1\over \sin\alpha}\sqrt{3Ic^3P\dot P\over 8\pi^2R^6},%
\label{B0}
\end{equation}
where $P=2\pi/\Omega$ is the rotation period, $I$ the moment of
inertia, $c$ the speed of light, $R$ the pulsar radius, $\alpha$
the inclination angle. $B$ is singular (i.e., $B\rightarrow
\infty$) when $\alpha=0^{\rm o}$. Therefore, the $B$-field
derivation in this way is questionable and inconsistent since
observation indicates $n<3$, which means other processes do
contribute to the braking torque.

Indeed some efforts have appeared to find unusual torque
mechanisms to understand the observed braking index (see, e.g.,
Menou, Perna \& Hernquist 2001, and references therein).
%
%
%
An alternative effort, within the framework of ``{\em standard}''
neutron stars and their magnetospheric emission models, is
proposed in this {\em Letter}. We find that $n$ and $B$ derivation
should generally depend on pulsar emission models. Assuming that
the orthogonal and aligned parts of magnetic moment are
responsible to the dipole-radiation torque and to the
unipolar-generator one, respectively, we obtain consistent
equations for calculating braking index and magnetic field in the
inner vacuum-gap, the space-charge-limited flow, and the outer gap
models. We find that all of these models result in braking index
$n<3$,
and in return the models could be tested for a particular pulsar
if its braking index and the inclination angle are observed.


\section{An assumption of the total energy loss for rotation-powered pulsars}

Pulsar broad-band emission depends essentially on a complete
solution of the formidable well-defined magnetosphere problem in
relativistic electrodynamics and plasma physics, which,
unfortunately, is still unknown hitherto (e.g., Mestel 2000).
Nevertheless, the problem has been understood to some extent in
two particular, i.e., the orthogonal and aligned rotating, cases.

\noindent {\bf Orthogonal Rotator}~~%
An orthogonal rotator with magnetic dipolar momentum $\mu_\perp$
emits monochromatic electromagnetic waves, the energy loss rate of
which is
$%
\dot E_{\rm d}=-{2\over 3c^3}\mu_{\bot}^2\Omega^4\simeq
-6.2\times 10^{27}~B_{12}^2~R_6^6~\Omega^4~{\rm ergs/s}%
$, %
where $B_{12}=B/(10^{12}{\rm G})$, $R_6=R/(10^6{\rm (cm)})$, and
$\mu_\perp=BR^3/2$.
%
These low frequency waves are generally unable to propagate and
should be absorbed in neutron star surroundings, and a larger
amount of energy and the corresponding momentum could be pumped
from neutron stars into their supernova remnants (Pacini 1967).
%
%

\noindent {\bf Aligned Rotator}~~%
%
The maximum potential drop in the open-field-line region by
unipolar effect is (e.g., Ruderman \& Sutherland 1975)
$%
\triangle\Phi={\mu_{\parallel}\Omega^2\over c^2}%
\simeq 5.56\times 10^8~B_{12}~R_6^3~\Omega^2~{\rm cgse}.%
$ %
$e^\pm$ pairs (or ions) are accelerated in charge depletion gaps,
picking up energy in the gaps and angular momentum from the
magnetic torque when streaming out. The angular momentum loss
requirement (Holloway 1977) can be satisfied if the charged
particles can be ``attached'' to the magnetic field as far as near
or out to the light cylinder. Two kinds of gaps are proposed to
work in pulsar magnetospheres, which are called as inner and outer
gaps. Various inner gaps are suggested, which depend on the
binding energy of charged particles in pulsar surface, e.g., the
vacuum gap model (Ruderman \& Sutherland 1975) with enough
binding, the space-charge-limited flow model without any binding
(Arons \& Scharlemenn 1979, Harding \& Muslimov 1998).
The outer gap model was suggested to work near the null surface
(e.g., Cheng, Ho \& Ruderman 1986, Zhang \& Cheng 1997) because
the charged particles on each side of the surface should flow in
opposite direction in order to close a global current in pulsar
magnetosphere.
It is thus obvious, as seen from above, that the energy loss is
model-dependent for aligned rotators, which will be considered
when calculating pulsar braking indices and magnetic fields in the
next section.
Nevertheless, the energy loss rate of an aligned rotator, due to
unipolar effect, could be written in the form of
$%
{\dot E_{\rm u}}=-2\pi r_{\rm p}^2\cdot c\varrho\cdot \Delta\phi%
$, %
if a gap has potential drop $\Delta\phi$ and the charge density in
the gap is $\varrho=\zeta \varrho_{\rm gj}\approx\zeta {\Omega
B\over 2\pi c}\simeq 5.3 \zeta B_{12}\Omega$ cgse cm$^{-3}$, where
the polar cap radius $r_{\rm p}=R\sqrt{R\Omega/c}\simeq 5.77\times
10^2R_6^{3/2}\Omega^{1/2}$ cm. $\zeta\sim 1$ since $\varrho$ and
$\varrho_{\rm gj}$ are conventionally expected to be in a same
order.

\noindent {\bf The Assumption}~~%
There are two schools of thought on the energy loss of an oblique
magnetized rotator. One group opined that the magnetodipole
radiation is the dominate mechanism of braking (e.g., Manchester
\& Taylor 1977, Dai \& Lu 1998, Lyubarsky \& Kirk 2001),
where no braking appears when $\alpha=0$.
Another group suggested that pulsars' spindown dominates by a
longitudinal current outflow due to unipolar generator (e.g.,
Beskin et al. 1984), where $\Omega=$ constant if $\alpha=90^{\rm
o}$.
However, although there are two unseemly points when $\alpha=0$
for the first school and when $\alpha=90^{\rm o}$ for the second
one, an interesting and strange thing, which is understandable in
the next section, is that the derived physical parameters (e.g.,
B-field strength) are reasonable.
We proposed that both energy loss mechanisms above, i.e., via
dipole radiation and via unipolar generator, are expected to
contribute the total braking torque of an oblique pulsar.
%
%
%
Phenomenologically, for a pulsar with a total magnetic momentum
${\vec \mu}={\vec \mu_\perp}+{\vec \mu_\parallel}$
($\mu_\perp=\mu\sin\alpha$, $\mu_\parallel=\mu\cos\alpha$), we
could write the total energy loss in the form of
%
$
{\dot E}=c_\perp {\dot E_{\rm d}}+c_\parallel {\dot E_{\rm u}}$, %
%
where $c_\perp$ and $c_\parallel$ are generally two functions of
$\alpha$ indicating the contributions of those two energy loss
mechanisms, respectively. Certainly $c_\perp(\alpha=\pi/2)=1$ and
$c_\parallel(\alpha=0)=1$. An essential and simple assumption to
be employed in this paper is $c_\perp = c_\parallel = 1$, since
$\dot E = \dot E_{\rm d} + \dot E_{\rm u}$ if $\mu_\perp$ and
$\mu_\parallel$ result {\em independently} in spindowns of $\dot
E_{\rm d}$ and $\dot E_{\rm u}$, respectively.
Therefore we have
\begin{equation}
{\dot E}=-{2\mu^2\over 3c^3}\Omega^4\eta,%
\label{total}
\end{equation}
with
$$
\begin{array}{lll}
\eta & \equiv & \sin^2\alpha+3\cos^2\alpha{\Delta\phi \over
\Delta\Phi}\\
& \simeq & \sin^2\alpha + 5.4 \times
10^{-9}R_6^{-3}B_{12}^{-1}\cos^2\alpha\Omega^{-2}\Delta\phi.
\end{array}
$$

\section{Braking index \& its implication}

The energy carried away by the dipole radiation ($\dot E_{\rm d}$)
and the relativistic particles ($\dot E_{\rm u}$) originates from
the rotation kinetic energy, the loss rate of which is $
\dot E=I\Omega\dot\Omega%
$. Energy conservation conduces towards
\begin{equation}
{\dot \Omega}=-{2\mu^2\over 3c^3I}\Omega^3\eta.%
\label{dotO}%
\end{equation}
Based on Eq.(\ref{dotO}), the braking index can be derived to be
\begin{equation}
n=3+{\Omega\dot\eta\over \dot\Omega\eta}=3+{\Omega\over \eta}
{{\rm d}\eta\over {\rm d}\Omega},%
\label{n}
\end{equation}
which is not exactly 3 as long as $\eta$ is not a constant. If
$\eta\propto \Omega^a$, then $n<3$ for $a<0$ ($n>3$ for $a>0$).
For pulsars near death line, $\Delta\phi \simeq \Delta\Phi$, i.e.,
the maximum potential drop $\Delta\Phi$ available acts on gap. In
this case, $\eta=1+2\cos^2\alpha<3$,
$\dot\eta=-2\sin(2\alpha)\dot\alpha$. $n<3$ if $\alpha$ gets
smaller as pulsar evolves.
For pulsars being away from death line, the potential drop
$\Delta\phi$ across an accelerator gap, which is model-dependent,
is much smaller than $\Delta\Phi$.
We discuss baking index in the following models, assuming that
$\vec{\mu}$ ($\mu$ and $\alpha$) and $I$ are not changed for
simplicity, since both observation (Bhattacharya et al. 1992) and
theory (e.g., Xu \& Busse 2001) imply that a pulsar's $B$-field
does not decay significantly during the rotation-powered phase.

\noindent
{\bf The vacuum gap (VG) model}~~ %
The basic picture of vacuum gap formed above polar cap with enough
binding energy was delineated explicitly in Ruderman \& Sutherland
(1975), where relativistic primary electrons emit $\gamma$-rays
via curvature radiation in the gap. The gap potential difference %
$\Delta\phi^{\rm VG}_{\rm CR}=4.1\times 10^9\rho_6^{4/7}
B_{12}^{-1/7}\Omega^{1/7}~{\rm cgse}$%
, where the curvature radius\footnote{%
Ruderman \& Sutherland (1975) supposed there are multipole
magnetic fields near pulsar surfaces, and they thus had
$\rho_6=1$. But in this paper we simply use dipole field lines for
indication.} %
$\rho=\rho_6 \times 10^6$ cm. $\rho\simeq {4\over
3}\sqrt{Rc/\Omega}\approx 2.3\times 10^8R_6^{1/2}\Omega^{-1/2}$
for polar cap accelerators. We thus have
$%
\Delta\phi^{\rm VG}_{\rm CR}=9.2\times 10^{10}R_6^{2/7}
B_{12}^{-1/7}\Omega^{-1/7}~{\rm cgse},~%
\eta^{\rm VG}_{\rm CR} \simeq \sin^2\alpha + 4.96 \times
10^2R_6^{-19/7}B_{12}^{-8/7}\cos^2\alpha\Omega^{-15/7}.
$ %
For vacuum gap where primary electrons emit $\gamma$-rays via
resonant inverse Compton scattering off the thermal photons (e.g.,
Zhang et al. 2000), the potential drop and the $\eta$ value are
$%
\Delta\phi^{\rm VG}_{\rm ICS}=1.9\times 10^{13}R_6^{4/7}
B_{12}^{-15/7}\Omega^{1/7}~{\rm cgse},~%
\eta^{\rm VG}_{\rm ICS} \simeq \sin^2\alpha + 1.02 \times
10^5R_6^{17/7}B_{12}^{-22/7}\cos^2\alpha\Omega^{-13/7}.
$%

\noindent
{\bf The space-charge-limited flow (SCLF) model}~~ %
SCLF model works for pulsars with boundary condition of
$E_\parallel=0$ at the pulsar surfaces.
The previous SCLF (Arons \& Scharlemann 1979) model has been
improved to a new version (e.g., Harding \& Muslimov 1998) with
the inclusion of the frame-dragging effect.
Though a simple and general analytical formula for all pulsar is
not available in the Harding-Muslimov (1998) model, the potential
drop could be well approximated in the extreme cases, regime I and
II, which are defined as cases without or with field
saturation\footnote{
The definitions of regime I and II in Zhang \& Harding (2000) have
been misprinted (B. Zhang, 2001, personal communication).
}. %
In regime II case (i.e., the gap height being larger than $r_{\rm
p}$), Zhang et al. (2000) obtained the potential drop, according
to which $\eta$ values can be calculated.
$%
\Delta\phi^{\rm SCLF}_{\rm II,CR}=7.1\times 10^9R_6^{3/4}
\Omega^{1/4}~{\rm cgse},~%
\eta^{\rm SCLF}_{\rm II,CR} \simeq \sin^2\alpha + 38
R_6^{-9/4}B_{12}^{-1}\cos^2\alpha\Omega^{-7/4},~
$%
for the CR-induced SCLF models;
$%
\Delta\phi^{\rm SCLF}_{\rm II,ICS}=4.2\times 10^8R_6^{28/13}
B_{12}^{-9/13}\Omega^{18/13}~{\rm cgse},~%
\eta^{\rm SCLF}_{\rm II,ICS} \simeq \sin^2\alpha + 2.3
R_6^{-11/13}B_{12}^{-22/13}\cos^2\alpha\Omega^{-8/13},~
$%
for the resonant ICS-induced SCLF models.
In regime I, the stable acceleration scenario should be controlled
by curvature radiation (Zhang \& Harding 2000),
$%
\Delta\phi^{\rm SCLF}_{\rm I}=1.8\times 10^{11}R_6^{4/7}
B_{12}^{-1/7}\Omega^{-1/7}~{\rm cgse},~%
\eta^{\rm SCLF}_{\rm I} \simeq \sin^2\alpha + 9.8
R_6^{-17/7}B_{12}^{-8/7}\cos^2\alpha\Omega^{-15/7}.~
$%

\noindent
{\bf The  outer gap (OG) model}~~ %
For a self-sustaining outer gap, which is limited by the $e^\pm$
pair produced by collisions between high-energy photons from the
gap and soft X-rays resulting from the surface heating by the
backflowing primary $e^\pm$ pairs, the potential drop is
$\Delta\phi=f^2\Delta\Phi$,
where the fractional size of such outer gap
$f=5.5B_{12}^{-4/7}P^{26/21}$ (Zhang \& Cheng 1997). $f<1$, which
is satisfied for the five pulsars, if outer gap works. The $\eta$
value therefore can be calculated,
$%
\Delta\phi^{\rm OG}=1.59\times 10^{12}R_6^3
B_{12}^{-1/7}\Omega^{-10/21}~{\rm cgse},~%
\eta^{\rm OG} \simeq \sin^2\alpha + 8.6\times 10^3
B_{12}^{-8/7}\cos^2\alpha\Omega^{-52/21}.
$%

From these $\eta$ values in different models, the braking index
can be obtained by Eq.(\ref{n}). For typical pulsars with $R_6=1$
and $B_{12}=1$, we compute the braking index $n$ in each model,
which is shown in Fig.\ref{f.n}. It is obvious that $n<3$ as long
as inclination angle $\alpha<90^{\rm o}$ in all of the models.
Pulsars with small rotation periods tend to have $n\approx 3$.
Also we can see from Fig.\ref{f.n} or Eq.(\ref{dotO}) that there
is a minimum braking index $n(\alpha=0^{\rm o})$ for each model.
In case of $B_{12}=R_6=1$, $n^{\rm VG}_{\rm CR}(\alpha=0^{\rm
o})=0.86$, $n^{\rm VG}_{\rm ICS}(\alpha=0^{\rm o})=1.14$, $n^{\rm
OG}(\alpha=0^{\rm o})=0.52$, $n^{\rm SCLF}_{\rm
II,CR}(\alpha=0^{\rm o})=1.25$, $n^{\rm SCLF}_{\rm
II,ICS}(\alpha=0^{\rm o})=2.38$, $n^{\rm SCLF}_{\rm
I}(\alpha=0^{\rm o})=0.86$.

We can not solve out magnetic field $B$ by only Eq.(\ref{dotO})
because $\eta=\eta(\alpha, \Omega)$. If $\alpha=90^{\rm o}$ (or
$\eta=1$), the solution of Eq.(\ref{dotO}) results in
Eq.(\ref{B0}). In principal, Eq.(\ref{dotO}) and (\ref{n}) should
be combined to find consistent $B$ and $\alpha$ in case of braking
index being known. However, because $1<\eta<3$, the magnetic field
derived from Eq.(\ref{B0}) is good enough but only modified by a
factor $1/\sqrt{\eta}\in(0.58,1)$.

Based on Eq.(\ref{dotO}) and (\ref{n}), the inclination angles of
the five pulsars with observed braking indices are calculated in
different models (see Table \ref{t.a}).
No solution of $\alpha$ is available for the Vela pulsar (PSR
B0833-45) and PSR B0540-69 for the regime II SCLF(ICS) model since
their braking indices are smaller than $n^{\rm SCLF}_{\rm
II,ICS}(\alpha=0^{\rm o})$. This is consistent with the fact that
these pulsars are young, and their gap heights are thus much
smaller than $r_{\rm p}$.


Furthermore, we can determine whether a model works on a
particular pulsar by comparing the calculated $\alpha$ in Table
\ref{t.a} with the observed $\alpha$.
Usually $\alpha$ can be derived by fitting the position angle
curves of pulsars with high linear polarization in the rotating
vector model (Lyne \& Manchester 1988).
For the five pulsars, only the inclination angle of the Vela
pulsar is obtained ($\sim 90^{\rm o}$), however no $\alpha$ value
in Table \ref{t.a} tallies with this observation. There may be two
possibilities 
%
%
to explain the discrepancy.
(1). The braking torques due to the dipole radiation and to the
unipolar generator should be treated and added in an other manner
(e.g., Harding et al. 1999), rather than the way of ours. However,
our treatment about the torques is reasonable, a further
improvement of braking calculation might not change substantially
the results presented.
(2). No model listed in Table \ref{t.a} can perfectly describe the
actual accelerate situation of the Vela pulsar. The outer gap
model explain well the high-energy emission of this pulsar, but
could be still a partial description of the global magnetosphere.
One possible picture is that both inner and outer gaps coexist in
a pulsar's magnetosphere (Usov 2000), but the {\em interaction}
between these two gaps and the pair plasma properties are still
very uncertain. It is also possible that pair production process
in strong magnetic and electric fields should be improved. For
example, if $B>0.1B_{\rm c}$ ($B_{\rm c}=4.4\times 10^{13}$G),
$\gamma$-photons nearly along curved field lines convert into
positroniums which could partially prevent the screening of
$E_\parallel$ (resultantly increasing the gap height and possibly
having $\zeta>1$), and therefore the energy loss $\dot E_{\rm u}$
increases significantly in polar cap models (Usov \& Melrose
1996). Such an increase could result in a larger $\alpha$ in Table
\ref{t.a} (see Eq.(\ref{n})) since all magnetic fields of the five
pulsar are very strong (near or greater than $0.1B_{\rm c}$).
In conclusion, further studies of testing emission models via
braking index and of the theoretical meaning of the test result
would be interesting and necessary.

\section{Conclusion \& Discussion}

We have proposed in this {\em Letter} that the observed braking
index $n<3$ could be understood if the braking torques due to the
dipole radiation and to the unipolar generator are combined.
The discrepancy between the observed inclination angle and that
derived from the six models of the Vela pulsar in Table \ref{t.a}
may call for improved pulsar emission models. In addition it is
found that the magnetic field strength of a pulsar by conventional
method could be a pretty good representation of the actual one.

%
%

\begin{figure}
\centerline{\psfig{figure=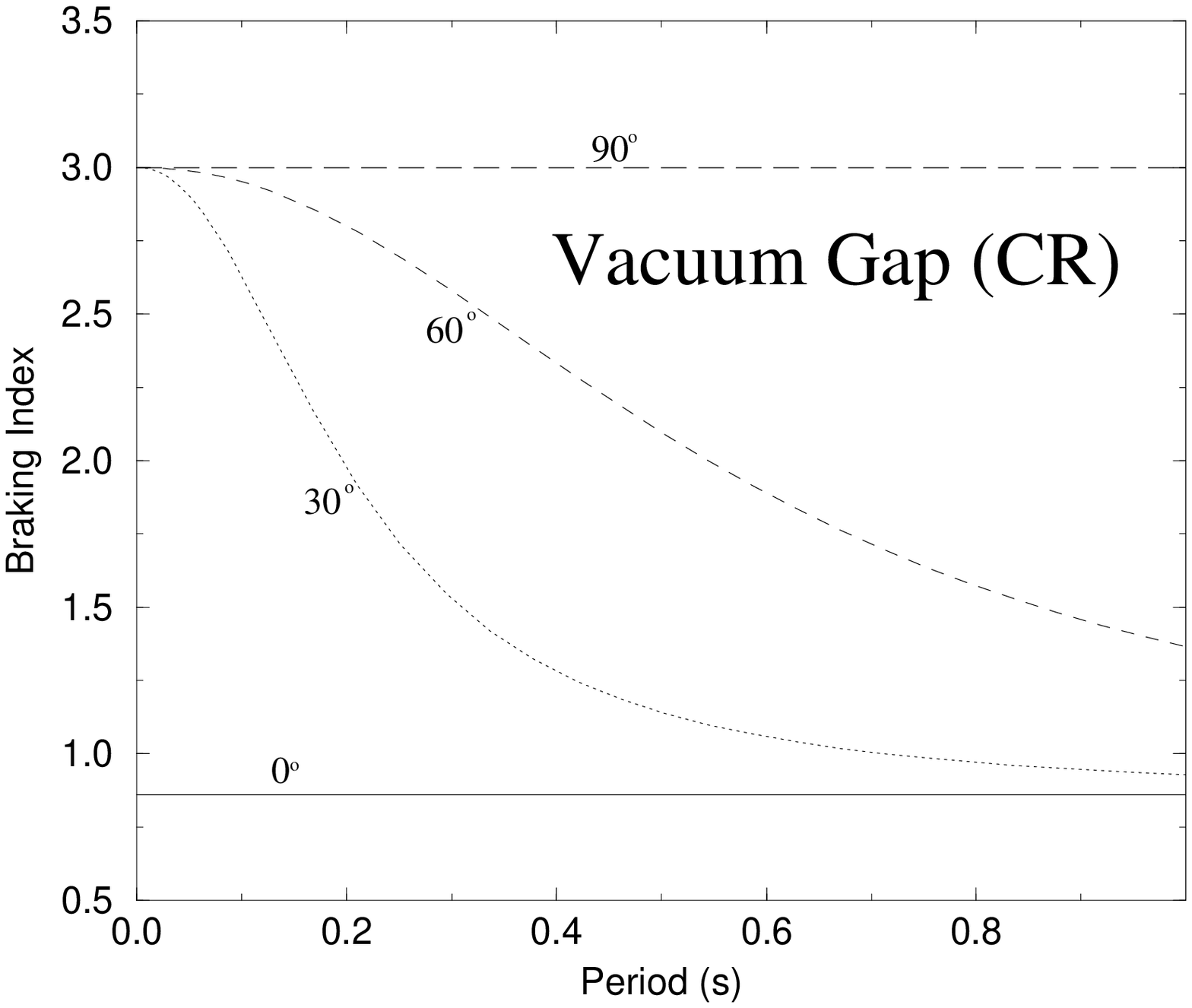,angle=0,height=6cm,width=6cm}\psfig{figure=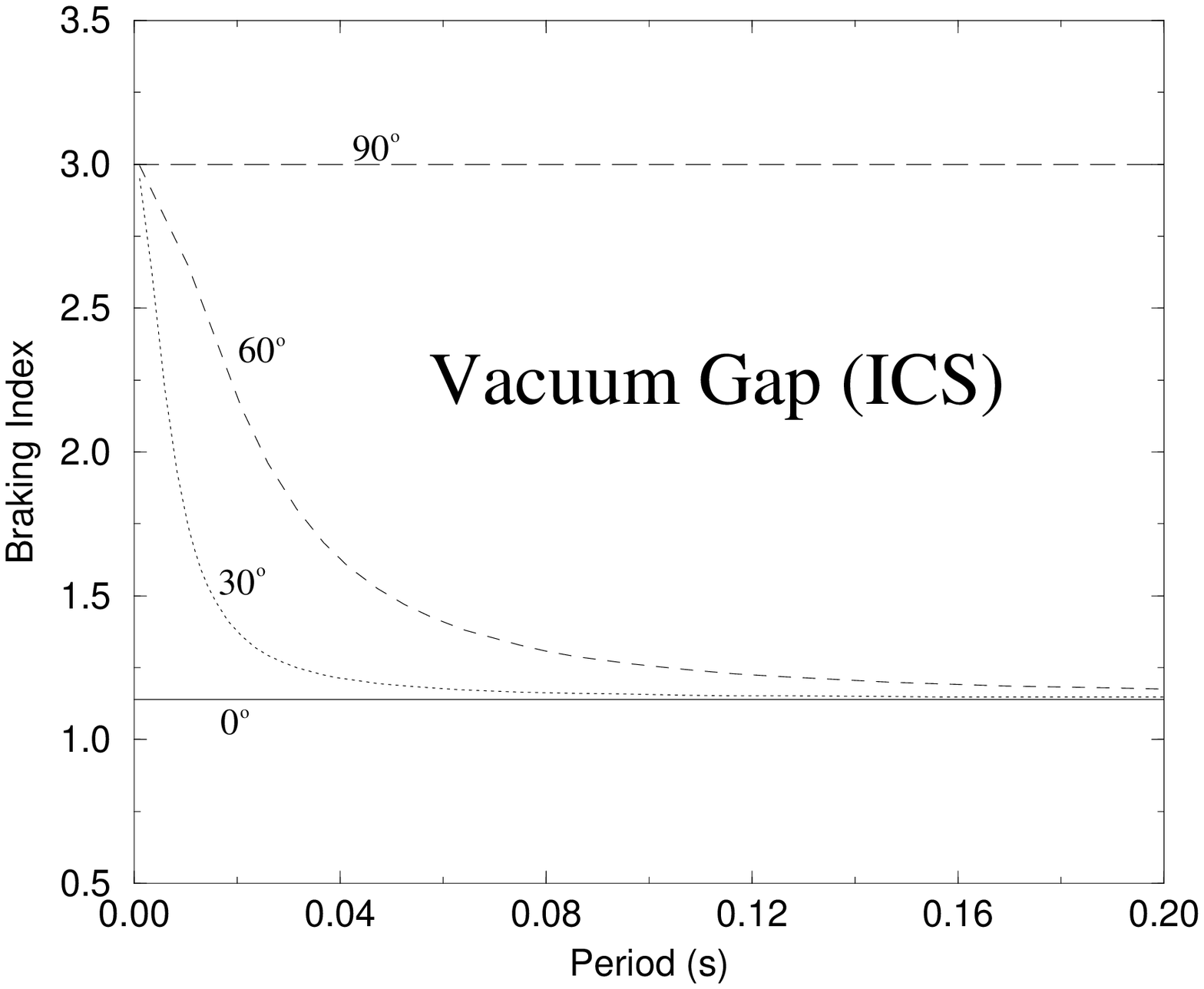,angle=0,height=6cm,width=6cm}\psfig{figure=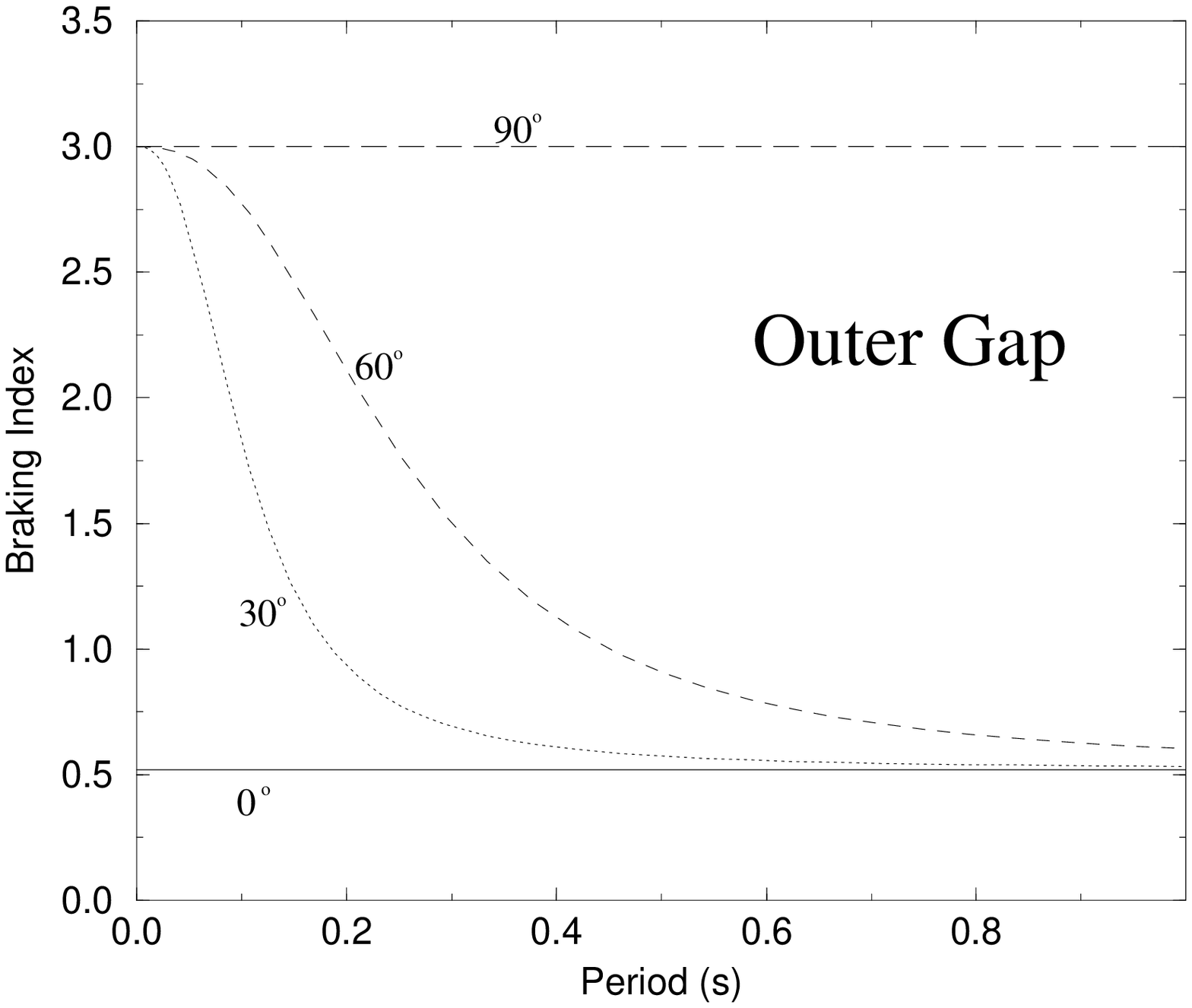,angle=0,height=6cm,width=6cm}}
\centerline{\psfig{figure=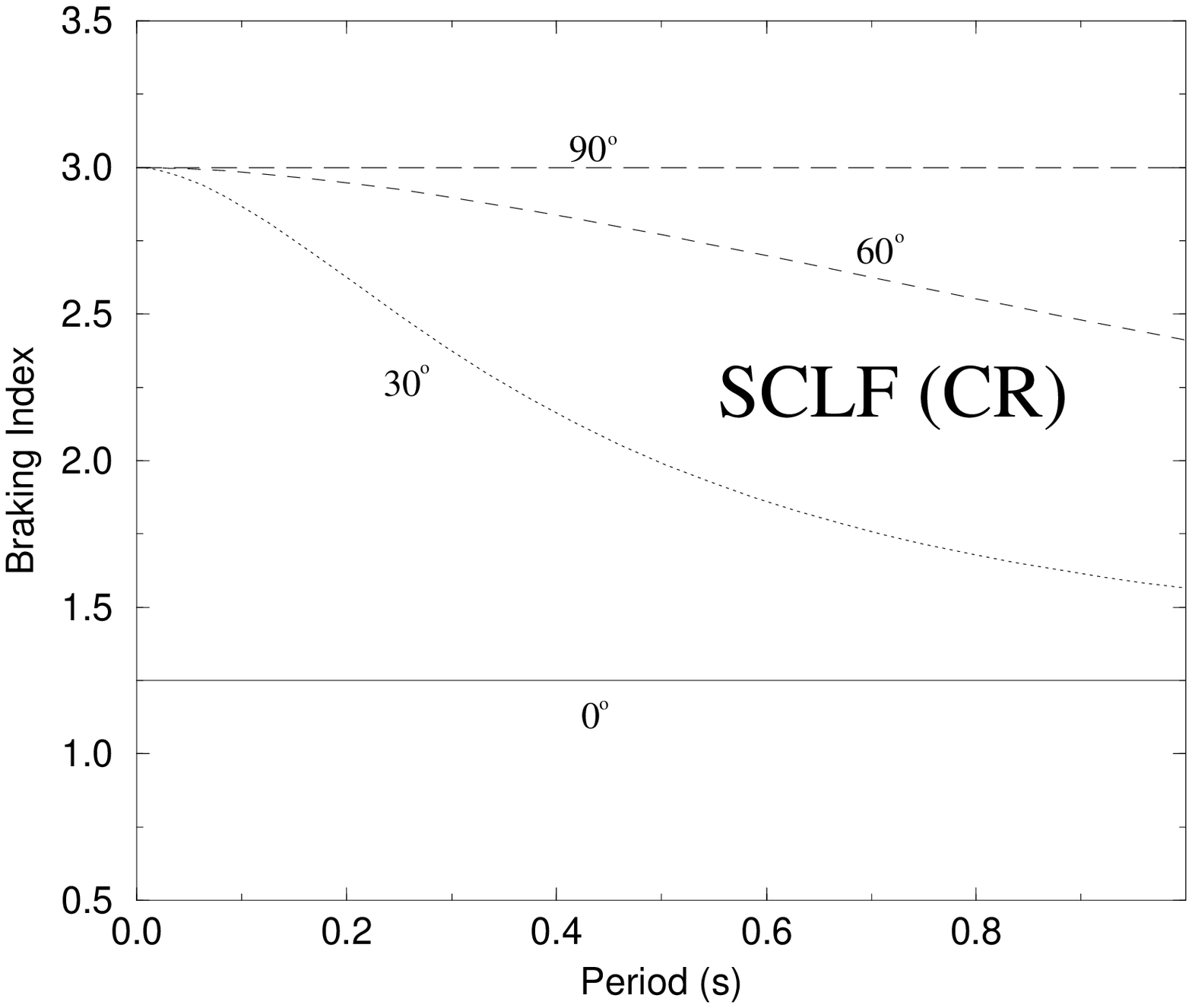,angle=0,height=6cm,width=6cm}\psfig{figure=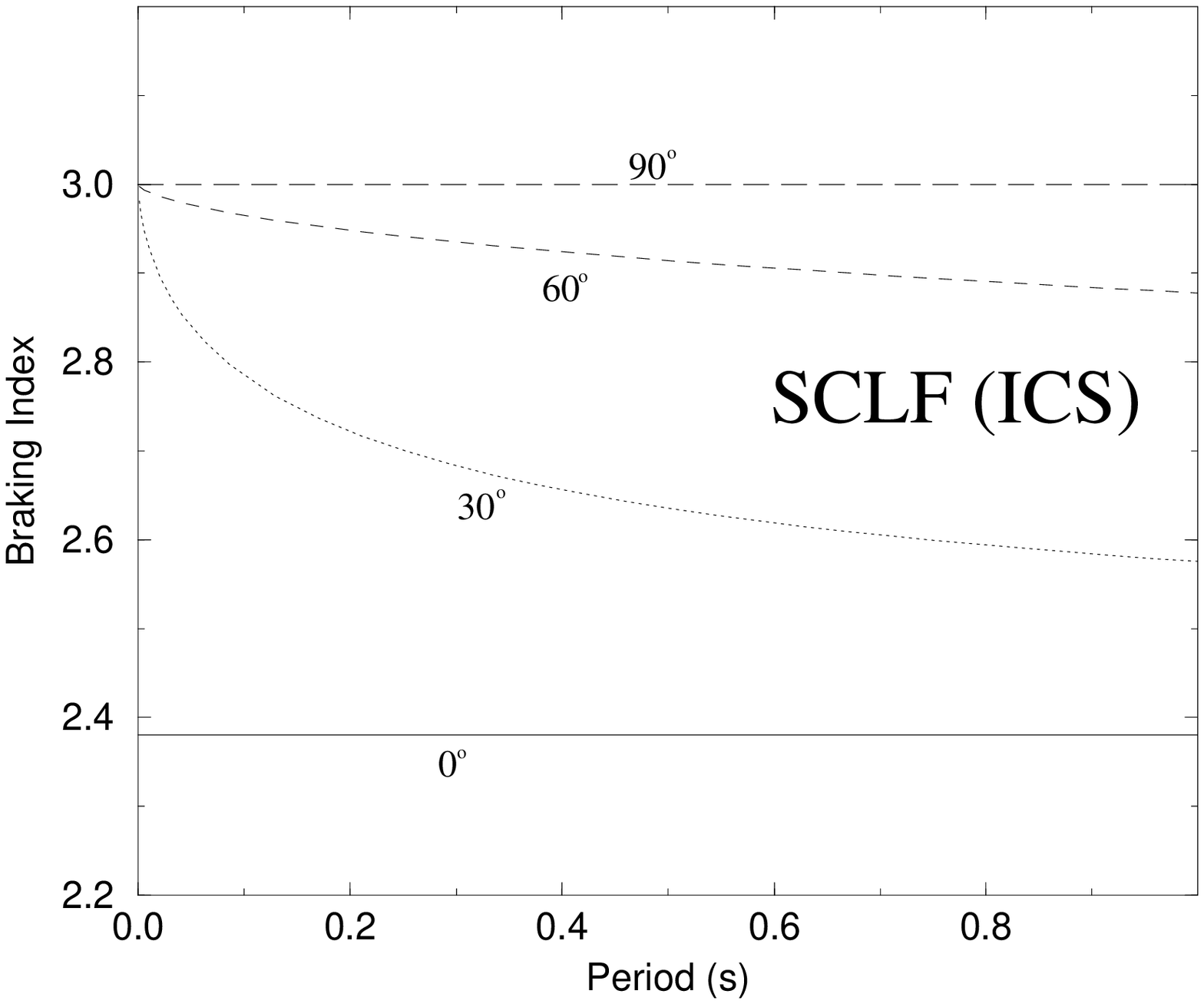,angle=0,height=6cm,width=6cm}\psfig{figure=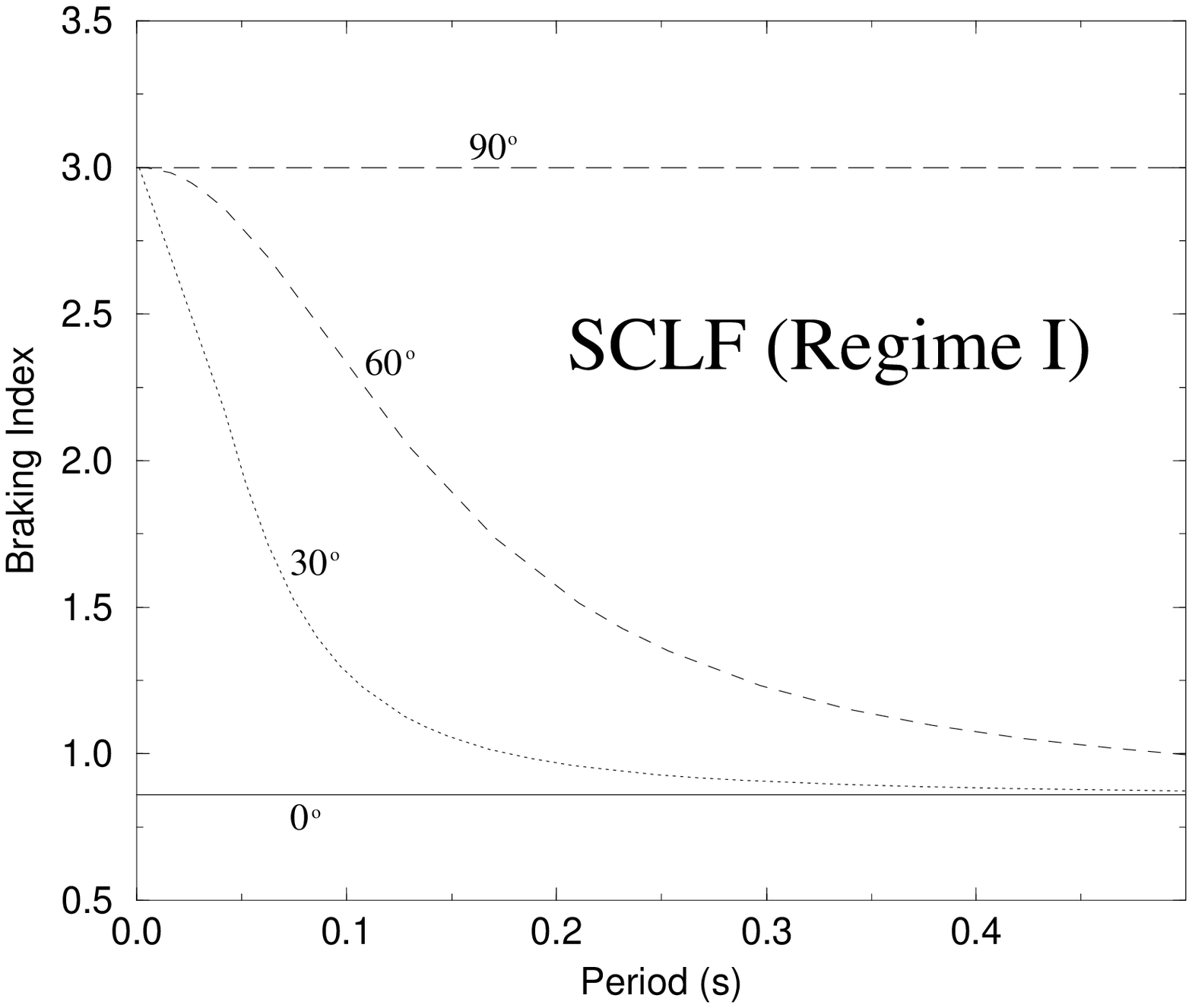,angle=0,height=6cm,width=6cm}}
\caption[]{%
A set of calculated braking indices, as functions of rotation
period, for six kinds of emission models. Pulsars are assumed to
have polar magnetic field $B=10^{12}$ G and radius $R=10^6$ cm
here. The inclination angles are chosen to be $0^{\rm o}$ (solid
lines), $30^{\rm o}$ (dotted lines), $60^{\rm o}$ (dashed lines),
and $90^{\rm o}$ (long-dashed lines). ``CR'' and ``ICS'' indicate
curvature-radiation-induced and resonant
inverse-Compton-scattering-induced gaps, respectively. SCLF(Regime
I): SCLF model without field saturation, SCLF(II,CR) and
SCLF(II,ICS): SCLF model with filed saturation (Regime II).
}%
\label{f.n}
\end{figure}

Fig.\ref{f.n} shows the variations of braking index $n$ as
functions of pulsar periods. Since pulsars spin down in their
life, the curves in Fig.\ref{f.n} represent the variations of $n$
as functions of pulsar ages to some extent. $n$ decreases as a
pulsar evolves.
However, the Johnston-Galloway's (1999) method to derive braking
index can only be applied if $n$ is constant during pulsar life.
Therefore $n$ can not been obtained by only $P$ and $\dot P$ in
principle.


\noindent {\it Acknowledgments}:
We sincerely thank Dr. Bing Zhang for his very helpful discussion,
comments and suggestions.
The authors wishes to acknowledge the nice computational and
academic resources of the Beijing Astrophysical Center.
This work is supported by National Nature Sciences Foundation of
China (19803001).

\begin{deluxetable}{lcccccc}
\tablewidth{6.5in} \tablenum{1}%
\tablecaption{The inclination angles ($\alpha$) of the five pulsars
derived from models}%
\tablehead{ \colhead{Name (PSR)} & \colhead{VG(CR)} &
\colhead{VG(ICS)} & \colhead{OG} & \colhead{SCLF(II,CR)} &
\colhead{SCLF(II,ICS)} & \colhead{SCLF(I)} }
\startdata%
B0531+21 & 2.6$^{\rm o}$ & 2.9$^{\rm o}$ & 5.0$^{\rm o}$
& 2.1$^{\rm o}$ & 1.6$^{\rm o}$ & 33$^{\rm o}$ \nl%
B0540-69 & 2.5$^{\rm o}$ & 6.3$^{\rm o}$ & 5.2$^{\rm o}$
& 1.8$^{\rm o}$ & --- & 36$^{\rm o}$ \nl%
B0833-45 & 2.6$^{\rm o}$ & 6.9$^{\rm o}$ & 6.7$^{\rm o}$
& 1.0$^{\rm o}$ & --- & 31$^{\rm o}$ \nl%
B1509-58 & 11$^{\rm o}$ & 8.4$^{\rm o}$ & 26$^{\rm o}$
& 7.6$^{\rm o}$ & 2.5$^{\rm o}$ & 81$^{\rm o}$ \nl%
J1119-6127 & 24$^{\rm o}$ & 6.3$^{\rm o}$ & 52$^{\rm o}$
& 15$^{\rm o}$ & 2.2$^{\rm o}$ & 88$^{\rm o}$ \nl%
\enddata
\label{t.a}
\end{deluxetable}{}
%

\end{document}